\documentstyle[12pt]{article}

\newcommand{\vx}{{\bf x}}
\newcommand{\vy}{{\bf y}}
\newcommand{\vv}{{\bf v}}
\newcommand{\vl}{{\bf \lambda}}
\newcommand{\vw}{{\bf w}}

\newcommand{\de}{\partial}
\newcommand{\ga}{\alpha}
\newcommand{\gb}{\beta}

\newcommand{\gl}{\lambda}
\title{Remark on the relation between passive scalars and diffusion backward
in time}
\author{Paolo Muratore-Ginanneschi}

\begin{document}
\maketitle
\centerline{N.B.I., Blegdamsvej 17 DK 2100, Copenhagen \O, Denmark}
\vfill
\begin{abstract}
The theory of stochastic differential equations is exploited to derive
the Hopf's identities for a passive scalar advected by a Gaussian drift 
field delta-correlated in time. The result holds true both for compressible 
and incompressible velocity field.
\end{abstract}
\vfill
\centerline{PACS: 05.20, 05.40, 47.20, 47.15}
\vfill
\pagebreak

\section{Introduction}
\label{s:intro}

In the last years much attention (see for example \cite{Krai, Kupi1,
Kupi2, Kupi3, Kupi4} and references therein) has been devoted to the study 
of passive scalar advected by an external field $\vv=\vv(\vx,t)$ 
and described by the equation
\begin{equation}
\de_{t} T+(\vv \cdot \nabla) T-\nu \Delta T=f 
\label{scal}
\end{equation}
where $\nu$ denotes the molecular diffusivity of the scalar $T=T(\vx,t)$ and
$f=f(\vx,t)$ is an external source driving the system.
The solution of (\ref{scal}) in a time interval $[0,t]$ is easily computed 
once it is known the kernel of the semigroup operator defined by the backward
Fokker-Planck equation
\begin{eqnarray}
\de_{t} P+ \nabla \cdot (\vv P) + \nu \Delta P=0 \nonumber \\
\lim_{s \uparrow t} P(\vy,s\,|\,\vx,t)=\delta(\vx-\vy)
\label{FP}
\end{eqnarray}
describing the conditional probability density for a particle to be in
position $\vy$ at time $s \leq t$ when it arrives in $\vx$ at time t.
 
The connection between the two equations is provided by the trajectories 
describing the realizations of a stochastic differential equation backward 
in time (BSDE), which represents the Lagrangian picture of the passive 
scalar equation (section \ref{s:lagra}). 
This is the  generalisation of the fact that in order to solve equation 
(\ref{scal}) for zero viscosity one has to invert the solutions 
of the ODE generated by the flow $\vv(\vx,t)$ .
 
The general solution of (\ref{FP}) can be written in the form of a path 
integral. This observation is very useful in the case of a random velocity 
(drift) field Gaussian in space and delta-correlated in time (section 
\ref{s:Krai}).     
There it is shown that for each N-points function of the theory i.e. 
for each average over the drift of the products of N fields at equal 
times, the Gaussian average defines a system of BSDEs of N variables. 
The equations satisfied by the N-points functions are the backward Kolmogorov 
equations corresponding to each of the BSDEs systems (section \ref{s:Hopf}).

\section{The Lagrangian picture}
\label{s:lagra}

Let us consider the stochastic differential equation
\begin{eqnarray}
d\vx_{s}&=&-\vv(\vx_{s},t-s)ds+\sqrt{2\,\nu} d\vw_{s} \quad s \leq t
\nonumber\\ 
\vx_{0}&=&\vx      
\label{sde} 
\end{eqnarray}
In general the solution of (\ref{sde}) exists and is unique if the drift 
field is Lipschitz and does not grow at infinity faster than $|\vx|^2$ 
(\cite{GS, kara}). It describes the motion of a particle localised at
time t in $\vx$ and diffusing for increasing $s$ i.e. backward in the global 
time $s'=t-s$
  
From the rules of Ito calculus the total differential of a scalar functional 
of (\ref{sde}) is given by:
\begin{eqnarray}
\lefteqn{d_{s} T(\vx_{s},t-s)=} \nonumber\\
&=&-[\de_{t-s} + \vv(\vx_{s},t-s) \cdot \nabla - \nu \Delta ]
T(\vx_{s},t-s) ds +\sqrt{2\,\nu} d\vw_{s} \cdot \nabla T(\vx_{s},t-s)      
\label{differential} 
\end{eqnarray}
The average of a Ito stochastic integral is zero therefore
if $T$ satisfies (\ref{scal}) with a given initial condition $T_{0}(\vx)$
at time $s=0$ its expression at time $t$ is provided by \cite{kara}:
\begin{equation}
T(\vx,t)=\int_{{\bf R}} d\,^{D}y T_{0}(\vy) P(\vy,0\,|\, \vx,t) +
\int_{0}^{t} ds \int_{{\bf R}} d\,^{D}y f(\vy,s) P(\vy,s\,|\, \vx,t)
\label{sol} 
\end{equation}
where $P$ is the transition probability density specified by (\ref{sde}):
\begin{equation}
P(\vy,s\,|\,\vx,t)=<\delta(\vy-\vx_{t-s})> 
\label{prob} 
\end{equation}
If one uses again Ito calculus
\begin{eqnarray}
&\,&\de_{s}P(\vy,s\,|\,\vx,t)=<-\,d_{t-s}\delta(\vy-\vx_{t-s})>= \nonumber\\
&=& <[\vv(\vx_{t-s},s) \cdot \nabla_{\vx_{t-s}}-\nu \Delta_{\vx_{t-s}}]
\delta(\vy-\vx_{t-s})]> 
\end{eqnarray}
by exploiting the properties of the delta function one arrives to 
(\ref{FP}) as mentioned in the introduction. 
Therefore the solution to the problem of the passive scalar is
equivalent to the investigation of the problem of backward diffusion. 
 
The transition probability (\ref{prob}) can be expressed in the form of a
path integral:
\begin{equation}
P(\vy,s\,|\,\vx,t)=({\cal S})\int_{\vx_{t-s}=\vy}^{\vx_{t}=\vx} 
{\cal D}x_{u}\,exp[-\int_{t-s}^{t} (\frac{\|\dot{\vx}_{u}+\,
\vv(\vx_{u},u)\|^{\,2}}{4\,\nu}-\frac{1}{2}\nabla\cdot \vv(\vx_{u},u))\,du]
\label{feystra} 
\end{equation}
The stochastic integral (\ref{feystra}) is evaluated according to the
Stratonovich's mid-point prescription as stressed by the symbol $({\cal S})$.
In the case of additive noise the Stratonovich prescription has the advantage 
to preserve under path integral sign the rules of ordinary calculus.

As it will become clear in the following, it is more useful for the scopes
of the present paper to resort to the Ito pre-point prescription which
allows to reabsorbe in the measure the divergence term in the short time
Lagrangian of (\ref{feystra}): 
\begin{equation}
P(\vy,s\,|\,\vx,t)=({\cal I})\int_{\vx_{t-s}=\vy}^{\vx_{t}=\vx} {\cal D}x_{u}\,
exp[-\int_{t-s}^{t} \frac{\|\dot{\vx}_{u}+\,\vv(\vx_{u},u)\|^{\,2}}{4\,\nu}\,
du]
\label{fey}
\end{equation}

\section{The Gaussian model}
\label{s:Krai}

Let us now assume that the velocity field in (\ref{scal}) is a Gaussian 
random field with zero average and correlation:
\begin{equation}
<v_{\ga}(\vx,t)v_{\gb}(\vy,s)>=\delta(t-s)D_{\ga\,\gb}(|\vx-\vy|)
\label{velcov} 
\end{equation}
Similar hypothesis are done for the external forcing:
\begin{equation}
<f(\vx,t)f(\vy,s)>=\delta(t-s)C(\vx-\vy)
\label{sourcov} 
\end{equation}
The assumption of time decorrelation is crucial in order to preserve 
under Gaussian average the locality in time of the short time action 
in (\ref{fey}).
On the other hand no assumption is done on the incompressibility of the 
velocity field in contrast with \cite{Kupi4}.

In order to compute the general N-points function one needs to evaluate
the path integral:
\begin{eqnarray}
\lefteqn{<\Pi_{l=1}^{N} P(\vy_{l},0 \,|\,\vx_{l},t_{l})>_{\vv}=}\nonumber\\
&=&\int{\cal D}\vv(\vx,u) \exp[-S(\vv)]\Pi_{l=1}^{N}\int_{\vx_{l}(0)=\vy_{l}}
^{\vx_{l}(t_{l})=\vx_{l}}{\cal D}\vx_{l}(u){\cal D}\vl_{l}(u)\,
exp[-S(\vx_{l},\vl_{l})]
\label{fey2} 
\end{eqnarray}
where
\begin{equation}
S(\vv)=\int_{0}^{t_{Max}}duds \int_{{\bf R}}d^{D}x
d^{D}y \, v_{\ga}(\vx,u)\frac{D_{\ga\,\gb}^{-1}(|\vx-\vy|)\delta(t-s)}{2}
v_{\gb}(\vy,s)
\end{equation}
and, after the Hubbart-Stratonovich transform:
\begin{eqnarray}
S(\vx_{l},\vl_{l})=\int_{0}^{t_{Max}}du\,\theta(t_{l}-u)[\,\nu \,|\vl_{l}(u)|
^{2}+i\,\vl_{l}(u)\cdot \dot{\vx}_{l}(u)+\,i\,\vl_{l}(u)\cdot 
\vv(\vx_{l}(u),u)]
\label{azione} 
\end{eqnarray}
In (\ref{azione}) $t_{Max}$ coincides with the largest $t_{l}$ and 
$\theta(t_{l}-u)$ is the usual step function.
The drift field appears linearly in (\ref{azione}): the Gaussian average 
can be easily performed. The averaged action is given by
\begin{eqnarray}
\lefteqn{S_{av}(\{\vx_{l}\}_{l=1}^{N},\{\vl_{l}\}_{l=1}^{N})=}\nonumber\\
&=&\Sigma_{l=1}^{N} \int_{0}^{t_{l}}du \{\gl_{l,\ga}(u)\frac{2\,\nu \delta_
{\ga\,\gb}+D_{\ga\,\gb}(0)}{2} \gl_{l,\gb}(u) + i\,\gl_{l,\ga}(u)
\dot{x}_{l,\ga}(u)\}+\nonumber\\
&+& \Sigma_{l=1, k>l}^{N} \int_{0}^{min(t_{l},t_{k})}du 
\gl_{l,\ga}(u)\frac{D_{\ga\,\gb}(|\vx_{l}(u)-\vx_{k}(u)|)}{2} \gl_{k,\gb}(u)
\label{effact}
\end{eqnarray}
where the Einstein convention and Greek letters have been used for vector 
indices ($\ga,\,\gb\,=\,1,...D$).
A further simplification is introduced by setting $t_{l}=t \,\forall\,l$:
\begin{eqnarray}
\lefteqn{S_{av}(\{\vx_{l}\}_{l=1}^{N},\{\vl_{l}\}_{l=1}^{N})=}
\nonumber\\
&=&\int_{0}^{t}du\,\{\gl_{l,\ga}(u)\frac{G_{\ga,\gb}(\vx_{l}(u)-\vx_{k}(u))}
{2}\gl_{l,\gb}(u)+ i\,\gl_{l,\ga}(u)\dot{x}_{l,\ga}(u)]
\label{eqtimes}
\end{eqnarray}
with the Einstein convention now extended to all indices and  
\begin{eqnarray}
&G_{\ga,\gb}(\vx_{l}-\vx_{k})&=G_{\ga,\gb}(0)=[2\,\nu \delta_{\ga\,\gb}+
D_{\ga\,\gb}(0)] \quad l=k \nonumber\\
&G_{\ga,\gb}(\vx_{l}-\vx_{k})&=D_{\ga\,\gb}(|\vx_{l}-\vx_{k}|) 
\quad l\neq k
\label{kernel}
\end{eqnarray}
From (\ref{eqtimes}), (\ref{kernel}) is immediately clear that the one point
function is Gaussian with constant diffusion matrix given by $G_{\ga,\gb}(0)$.
On the other hand non-trivial behaviour is expected for $N \geq 2$.
By integrating over the ghost trajectories $\{\vl_{l}\}_{l=1}^{N}$ 
one obtains
\begin{equation}
S_{av}(\{\vx_{l}\}_{l=1}^{N})=\int_{0}^{t}du \,\dot{x}_{l,\ga}(u)
\frac{G_{\ga,\gb}^{-1} (\vx_{l}(u)-\vx_{k}(u))}{2} \dot{x}_{l,\gb}(u)
\label{Itores}
\end{equation}
The averaged action (\ref{Itores}) is exactly the one that we would have
found starting from the system of Ito stochastic differential equations (SDE):
\begin{equation}
dx_{l,\ga}(t)=G_{\ga,\gb}^{\frac{1}{2}}(\vx_{l}(t)-\vx_{k}(t))dw_{k,\gb}(t)
\quad l,k=1,N
\label{stosys}
\end{equation}
where $dw(t)$ is the Wiener differential.
For N=1 it is again evident that the diffusion is normal with 
diffusion matrix $G_{\ga,\gb}(0)$. 
The effect of the Gaussian average is therefore to associate to each of 
the N points function an effective diffusion.

Two remarks are needed.
First, the system (\ref{stosys}) is autonomous. This implies   
that the relation between the transition probability density for the forward 
process and that one of the backward process is particularly simple:
\begin{eqnarray}
{\cal P}_{f}^{(N)}(\{\vx_{l}\}_{l=1}^{N},t\,|\{\vy_{l}\}_{l=1}^{N},s)&=&
{\cal P}_{b}^{(N)}(\{\vx_{l}\}_{l=1}^{N},s\,|\{\vy_{l}\}_{l=1}^{N},t)
\nonumber\\
&=&{\cal F}^{(N)}(\{\vx_{l}\}_{l=1}^{N},\{\vy_{l}\}_{l=1}^{N},t-s)
\label{equiv}
\end{eqnarray}
for any $t \geq s$ since that 
\begin{eqnarray}
\de_{t}{\cal F}^{(N)}=-\de_{s}{\cal F}^{(N)}&=&\frac{1}{2}\de_{l,\ga}
\de_{k,\ga}[G_{\ga,\gb}^{l,k}{\cal F}^{(N)}]\nonumber\\
{\cal F}^{(N)}(\{\vx_{l}\}_{l=1}^{N},\{\vy_{l}\}_{l=1}^{N},0)&=&
\Pi_{l=1}^{N}\delta(\vx_{l}-\vy_{l})
\end{eqnarray}
Second, had we used the Stratonovich prescription we would have got to 
the averaged action:
\begin{eqnarray}
\lefteqn{S_{av}(\{\vx_{l}\}_{l=1}^{N})=}\nonumber\\
&=&\int_{0}^{t}du\,[ \dot{x}_{l,\ga}(u) 
\frac{G_{\ga,\gb}^{-1}(\vx_{l}(u)-\vx_{k}(u))}{2} \dot{x}_{k,\gb}(u)-
\frac{1}{4} \de_{l,\ga} \de_{k,\gb} D_{\ga\,\gb}(|\vx_{l}(u)-
\vx_{k}(u)|)]
\label{Strares}
\end{eqnarray}
Whilst its probabilistic meaning is a posteriori clear, (\ref{Strares}) has 
the disadvantage of a less intuitive form.

\section{The Hopf's identities}
\label{s:Hopf}

Let us define the average of a given function $F_{0}(\vx,...,\vx)$ 
over the realizations of (\ref{stosys}) interpreted as BSDEs:
\begin{equation}
<F(\vx_{1}(t),...,\vx_{N}(t))>=\Pi_{l=1}^{N}\int_{-\infty}^{\infty}
d^{D}y_{l} F_{0}(\vx,...,\vx) {\cal P}_{b}^{(N)}(\vy_{1},...,\vy_{N},s\,
|\,\vx_{1},...,\vx_{N},t)
\end{equation}
Then \cite{kara}
\begin{equation}
\de_{t}<F(\vx_{1}(t),...,\vx_{N}(t))>=\frac{1}{2}G_{\ga,\gb}(\vx_{l}
-\vx_{k})\de_{l,\ga} \de_{k,\gb} <F(\vx^{1}_{t},...,\vx^{N}_{t})>
\label{formula}
\end{equation}
where $\frac{1}{2}G_{\ga,\gb}(\vx_{l}-\vx_{k})\de_{l,\ga}\de_{k,\gb}$ is 
equal to the operator $-\,{\cal M_{N}}$ of \cite{Kupi1, Kupi2, Kupi3, Kupi4} 
Equation (\ref{formula}) is the backward Kolmogorov equation for the
system (\ref{stosys}) interpreted as BSDEs so it is perfectly consistent 
with initial conditions in time.

The Hopf's identities are a straightforward consequence of this observation.
Namely starting from (\ref{sol}), let us consider the case of the two 
point functions.
\begin{eqnarray}
\lefteqn{\de_{t}<<T(\vx_{1},t)\,T(\vx_{2},t)>_{f}>_{\vv}=}\nonumber\\ 
&=&\int_{0}^{t}ds\,\int_{-\infty}^{\infty}d^{D}y_{1}d^{D}y_{2}[T_{0}(\vy_{1})
T_{0}(\vy_{2})\delta(s)+C(\vy_{1}-\vy_{2})\,]\de_{t}
{\cal P}_{b}^{(2)}(\vy_{1},\vy_{2},s\,|\,\vx_{1},\vx_{2},t)+\nonumber\\
&+&\int_{-\infty}^{\infty}d^{D}y_{1}d^{D}y_{2}\,C(\vy_{1}-\vy_{2})\,
{\cal P}_{b}^{(2)}(\vy_{1},\vy_{2},t\,|\,\vx_{1},\vx_{2},t)
\label{1step}
\end{eqnarray}
where $<...>_{f}$ denotes the average over the external force.
By means of (\ref{formula}), we recognise that (\ref{1step}) can be 
rewritten as:
\begin{equation}
\de_{t} <<T(\vx_{1},t)\,T(\vx_{2},t)>_{f}>_{\vv}=-{\cal M}_{2}
<<T(\vx_{1},t)\,T(\vx_{2},t)>_{f}>_{\vv}+C(\vx_{1}-\vx_{2})
\label{2pti}
\end{equation}
In the limit $t \uparrow \infty$ (stationary case) we obtain the Hopf's 
identity for the two point function. 

Let us turn to the four point function. 
\begin{eqnarray}
\lefteqn{\de_{t}<<\Pi_{l=1}^{4}T(\vx_{l},t)>_{f}>_{\vv}>=}
\nonumber\\
&=&\int_{0}^{t}ds\,du \int_{-\infty}^{\infty}\Pi_{l=1}^{4}d^{D}y_{l}
\,[\Pi_{l=1}^{4}T_{0}(\vy_{l})\delta(s)\delta(s-u)+T_{0}(\vy_{1})
T_{0}(\vy_{2})\delta(s)C(\vy_{3}-\vy_{4})\,+
\nonumber\\
&+&C(\vy_{1}-\vy_{2})C(\vy_{3}-\vy_{4})\,]\,
\de_{t}{\cal P}_{b}^{(4)}(\{\vy_{1},\vy_{2},s\}\{\vy_{3},\vy_{4},u\}\,|\,
\vx_{1},...,\vx_{4},t)+perm.+
\nonumber\\
&+&C(\vx_{1}-\vx_{2})\,\int_{0}^{t}ds\int_{-\infty}^{\infty}
d^{D}y_{3}d^{D}y_{4}\,[T_{0}(\vy_{3})T_{0}(\vy_{4})\delta(s)+
\nonumber\\
&+&C(\vy_{3}-\vy_{4})]\,{\cal P}_{b}^{(2)}(\vy_{3},\vy_{4},s\,
|\,\vx_{3},\vx_{4},t)+ \nonumber\\
&+& perm. 
\label{2step}
\end{eqnarray}
here $perm.$ means permutation among the $l-indixes$. 

The expression 
${\cal P}_{b}^{(4)}(\{\vy_{1},\vy_{2},s\}\{\vy_{3},\vy_{4},u\}\,|\,
\vx_{1},...,\vx_{4},t)$ denotes the probability density conditioned 
upon an event at later time $t$ that the components for $l=1,2$ of the 
system (\ref{stosys}) take the value $\vy_{1},\vy_{2}$ at a prior time
$s$ whilst those for $l=3,4$ the value $\vy_{3},\vy_{4}$ at time $u \leq t$.
Since we are interpreting (\ref{stosys}) as BSDEs with final conditions
this event is perfectly well defined.
Therefore we conclude:
\begin{eqnarray}
\lefteqn{\de_{t} <<\Pi_{l=1}^{4}T(\vx_{l},t)>_{f}>_{\vv}>=}
\nonumber\\
&=&-{\cal M}_{4}<<\Pi_{l=1}^{4}T(\vx_{l},t)>_{f}>_{\vv}>+
C(\vx_{1}-\vx_{2})<<T(\vx_{3},t)\,T(\vx_{4},t)>_{f}>_{\vv}+\nonumber\\
&+&perm.
\label{4pti}
\end{eqnarray}
In the limit $t \uparrow \infty$ we obtain:
\begin{eqnarray}
\lefteqn{{\cal M}_{4}<<T(\vx_{1})\,T(\vx_{2})\,T(\vx_{3})\,T(\vx_{4})>_{f}>
_{\vv}=}\nonumber\\
&=&C(\vx_{1}-\vx_{2})<<T(\vx_{3})\,T(\vx_{4})>_{f}>_{\vv}+ 
permutations
\end{eqnarray}

The Hopf's identity satisfied by the general $N$-points correlation function 
can be derived in the same way.

\section{Conclusion}
\label{s:Con}

By means of the theory of stochastic differential equations the Hopf's 
identities have been proven for a passive scalar advected by 
a Gaussian delta-correlated in time velocity field for both the compressible 
and incompressible case.

\section{Acknowledgements}

I warmly thanks E. Aurell and A. Vulpiani for their precious teaching and 
constant encouragement.
I am also pleased to thank M. H. Jensen, P. Olesen and H. Fogedby for 
comments and useful suggestions. 

This work was supported by a TMR grant (ERB4001GT962476) from the European 
Commission.


\begin{thebibliography}{99}


\bibitem{Krai} 
R.H. Kraichnan, Phys. Rev. Lett. {\bf 72} (1994), 1016-1019.
\bibitem{Kupi1}
K. Gawedzki and A. Kupiainen, Phys. Rev. Lett. {\bf 75} (1995), 3834-3837.
\bibitem{Kupi2}
K. Gawedzki and A. Kupiainen, chao-dyn/9504002
\bibitem{Kupi3}
K. Gawedzki and A. Kupiainen, chao-dyn/9601018 
\bibitem{Kupi4}
D. Bernard K. Gawedzki and A. Kupiainen, cond-mat/9706035
\bibitem{GS}
I.J. Gihman and A.V. Skorohod (1972) {\it Stochastic Differential Equations}
\\ Springer-Verlag Berlin.
\bibitem{kara}
I. Karatzas and S.E. Shreve (1991) {\it Brownian Motion and Stochastic 
Calculus}\\ (2nd Edition) Springer-Verlag Berlin.
\end{thebibliography}
\end{document}